\documentclass{aa}

\usepackage{lscape}
\usepackage{graphicx}
\usepackage{txfonts}
\usepackage{multirow}
\usepackage{hyperref}
\begin{document}

    \titlerunning{Photometry and spectroscopy of the Type~Icn SN~2021ckj.}  
    \authorrunning{T. Nagao et al.}

   
   \title{Photometry and spectroscopy of the Type Icn supernova 2021ckj}
   \subtitle{The diverse properties of the ejecta and circumstellar matter of Type Icn SNe}

   \author{T.~Nagao,\inst{1,2,3}\fnmsep\thanks{takashi.nagao@utu.fi} 
          H.~Kuncarayakti,\inst{1,4} 
          K.~Maeda,\inst{5} 
          T.~Moore,\inst{6} 
          A.~Pastorello,\inst{7} 
          S.~Mattila,\inst{1,8} 
          K.~Uno,\inst{5} 
          S.~J.~Smartt,\inst{6} 
          S.~A.~Sim, \inst{6} 
          L.~Ferrari,\inst{9,10} %
          L.~Tomasella,\inst{7} 
          J.~P.~Anderson,\inst{11,12} 
          T.-W. Chen,\inst{13,14} 
          L.~Galbany, \inst{15,16} 
          H.~Gao,\inst{17}
          M.~Gromadzki,\inst{18} 
          C.~P.~Guti{\'e}rrez,\inst{4,1} 
          C.~Inserra,\inst{19} 
          E.~Kankare,\inst{1,20} 
          E. A. Magnier,\inst{21} %
          T.~E.~M\"uller-Bravo,\inst{15,16} 
          A.~Reguitti, \inst{22,12,7} 
          \and
          D.~R.~Young\inst{6} 
          }

   \institute{Department of Physics and Astronomy, University of Turku, FI-20014 Turku, Finland
         \and
         Aalto University Mets\"ahovi Radio Observatory, Mets\"ahovintie 114, 02540 Kylm\"al\"a, Finland
         \and
         Aalto University Department of Electronics and Nanoengineering, P.O. BOX 15500, FI-00076 AALTO, Finland
         \and
         Finnish Centre for Astronomy with ESO (FINCA), FI-20014 University of Turku, Finland
         \and
         Department of astronomy, Kyoto University, Kitashirakawa-Oiwake-cho, Sakyo-ku, Kyoto 606-8502, Japan
         \and
         Astrophysics Research Centre, School of Mathematics and Physics, Queens University Belfast, Belfast BT7 1NN, UK
         \and
         INAF – Osservatorio Astronomico di Padova, Vicolo dell'Osservatorio 5, 35122 Padova, Italy
         \and
         School of Sciences, European University Cyprus, Diogenes street, Engomi, 1516 Nicosia, Cyprus
         \and
         Facultad de Ciencias Astronómicas y Geofísicas, Universidad Nacional de La Plata, Paseo del Bosque S/N, B1900FWA La Plata, Argentina
         \and
         Instituto de Astrofísica de La Plata (IALP), CONICET, Argentina
         \and
         European Southern Observatory, Alonso de C\'ordova 3107, Casilla 19, Santiago, Chile
         \and
         Millennium Institute of Astrophysics MAS, Nuncio Monsenor Sotero Sanz 100, Off. 104, Providencia, Santiago, Chile
         \and
         Technische Universit{\"a}t M{\"u}nchen, TUM School of Natural Sciences, Physik-Department, James-Franck-Stra{\ss}e 1, 85748 Garching, Germany
         \and
         Max-Planck-Institut f{\"u}r Astrophysik, Karl-Schwarzschild Stra{\ss}e 1, 85748 Garching, Germany
         \and
         Institute of Space Sciences (ICE, CSIC), Campus UAB, Carrer de Can Magrans, s/n, E-08193 Barcelona, Spain
         \and
         Institut d’Estudis Espacials de Catalunya (IEEC), E-08034 Barcelona, Spain
         \and
         Institute for Astronomy, University of Hawaii, 2680 Woodlawn Drive, Honolulu HI 96822, USA
         \and
         Astronomical Observatory, University of Warsaw, Al. Ujazdowskie 4, 00-478 Warszawa, Poland
         \and
         Cardiff Hub for Astrophysics Research and Technology, School of Physics \& Astronomy, Cardiff University, Queens Buildings, The Parade, Cardiff, CF24 3AA, UK
         \and
         Turku Collegium for Science, Medicine and Technology, University of Turku, FI-20014 Turku, Finland
         \and
         Institute of Astronomy, University of Hawaii, 2680 Woodlawn Drive, Honolulu, HI 96822,  USA
         \and
         Instituto de Astrof\'{i}sica, Departamento de F\'{i}sica – Universidad Andres Bello, Avda. Rep\'{u}blica 252, 8320000 Santiago, Chile
             }

   \date{Received ?? ??, 2023; accepted ?? ??, 2023}

 
  \abstract
   {We present photometric and spectroscopic observations of the Type Icn supernova (SN) 2021ckj. This rare type of SNe is characterized by a rapid evolution and high peak luminosity as well as narrow lines of highly-ionized carbon at early phases, implying interaction with hydrogen- and helium-poor circumstellar-matter (CSM). SN~2021ckj reached a peak brightness of $\sim -20$ mag in the optical bands, with a rise time and a time above half-maximum of $\sim 4$ and $\sim 10$ days, respectively, in the $g$/$cyan$ bands. These features are reminiscent of those of other Type~Icn SNe (SNe~2019hgp, 2021csp and 2019jc), with the photometric properties of SN~2021ckj being almost identical to those of SN~2021csp. 
   Spectral modeling of SN~2021ckj reveals that its composition is dominated by oxygen, carbon and iron group elements, and the photospheric velocity at peak is $\sim 10000$ km s$^{-1}$. Modeling of the spectral time series of SN~2021ckj suggests aspherical SN ejecta.
   From the light curve (LC) modeling applied to SNe 2021ckj, 2019hgp, and 2021csp, we find that the ejecta and CSM properties of Type Icn SNe are diverse.
   SNe~2021ckj and 2021csp likely have two ejecta components (an aspherical high-energy component and a spherical standard-energy component) with a roughly spherical CSM, while SN~2019hgp can be explained by a spherical ejecta-CSM interaction alone. The ejecta of SNe~2021ckj and 2021csp have larger energy per ejecta mass than the ejecta of SN~2019hgp.
   The density distribution of the CSM is similar in these three SNe, and is comparable to those of Type~Ibn SNe. This may imply that the mass-loss mechanism is common between Type~Icn (and also Type~Ibn) SNe. The CSM masses of SN~2021ckj and SN~2021csp are higher than that of SN~2019hgp, although all these values are within the diversity seen in Type~Ibn SNe.
   The early spectrum of SN~2021ckj shows narrow emission lines from C~II and C~III, without a clear absorption component, in contrast with that observed in SN 2021csp. The similarity of the emission components of these lines implies that the emitting regions of SNe~2021ckj and 2021csp have similar ionization states, and thus suggests that they have similar properties of the ejecta and CSM, which is inferred also from the LC modeling. Taking into account the difference in the strength of the absorption features, this heterogeneity may be attributed to viewing angle effects in otherwise common aspherical ejecta. In particular, in this scenario SN~2021ckj is observed from the polar direction, while SN~2021csp is seen from an off-axis direction.
   This is also supported by the fact that the late-time spectra of SNe~2021ckj and 2021csp show similar features but with different line velocities.
   }


   \maketitle
%

\section{Introduction}

   A new class of supernovae (SNe) has been recently proposed based on the observations of SN~2019hgp \citep[Type Icn SNe;][]{Gal-Yam2022}. These SNe are characterized by a rapid photometric evolution ($t_{\rm{rise}} \lesssim 10$ days) with a high peak brightness ($r\sim -18.5$ mag) as well as narrow P-Cygni lines of highly-ionized carbon, oxygen and neon in its early spectra \citep[][]{Gal-Yam2022}. Recently, several more examples, which show similar observational properties, have been reported \citep[SNe~2019jc, 2021csp, 2021ckj and 2022ann;][]{Perley2022, Fraser2021, Pellegrino2022, Davis2022}. The narrow lines of such highly-ionized elements suggest an interaction of the SN ejecta with dense hydrogen- and helium-poor circumstellar matter (CSM).

   \citet[][]{Fraser2021} estimated that the bolometric light curve (LC) of SN~2021csp can be reproduced by a $4\times 10^{51}$ erg explosion with 2 M$_{\odot}$ of ejecta and 0.4 M$_{\odot}$ of $^{56}$Ni, plus a contribution from shock cooling emission of $\sim 1$ M$_{\odot}$ of CSM extending out to 400 R$_{\odot}$.
   However, from their LC analysis of SN~2021csp, \citet[][]{Perley2022} and \citet[][]{Pellegrino2022} favor CSM interaction as the main energy source of this SN.
   \citet[][]{Gal-Yam2022} also demonstrated that the bolometric light curve of SN~2019hgp is well fit by a CSM interaction model (progenitor radius of $4.1 \times 10^{11}$ cm, ejecta mass of 1.2 M$_{\odot}$, opacity of 0.04 cm$^{2}$ g$^{-1}$, CSM mass of 0.2 M$_{\odot}$, a mass loss rate of 0.004 M$_{\odot}$ yr$^{-1}$, and an expansion speed of 1900 km s$^{-1}$) rather than models with an energy input from the radioactive decay of $^{56}$Ni/$^{56}$Co.

   Since Type Icn SNe do not show hydrogen or helium features in their spectra, their progenitors are believed to be stars whose hydrogen/helium envelopes were stripped, as inferred for the progenitors of classical Type~Ic SNe. Therefore, a straightforward scenario for the origin of Type~Icn SNe would be a similar progenitor star as those of classical Type~Ic SNe, exploding just after an extensive mass ejection, even though the progenitors of classical Type~Ic SNe are also under debate \citep[see, e.g.,][and references therein]{Yoon2015}.
   However, \citet[][]{Perley2022} estimated a minimal $^{56}$Ni mass and/or a low ejecta mass from their late-time deep photometry for SN~2021csp, and concluded that its progenitor is intrinsically different from those of classical Type Ic SNe. This conclusion is also supported by \citet[][]{Pellegrino2022}, who inferred low ejecta masses ($\lesssim$ 2 M$_{\odot}$) and low $^{56}$Ni masses ($\lesssim$ 0.04 M$_{\odot}$) from the LCs of four Type~Icn SNe. These values are lower than typical $^{56}$Ni masses ($\sim 0.2$ M$_\odot$) and ejecta masses ($\sim 2$ M$_\odot$) estimated for Type~Ic SNe \citep[e.g.,][]{Drout2011}.

   There are several proposed scenarios for the origin of Type Icn SNe \citep[e.g.,][]{Perley2022, Fraser2021}.
   For example:
   (1) an SN from a highly stripped star in a binary system \citep[e.g.,][]{De2018, Sawada2022};
   (2) a Pulsational Pair-Instability SN (PPISN) \citep[e.g.,][]{Woosley2017};
   (3) a merger of a Wolf-Rayet (WR) star and a compact object \citep[e.g.,][]{Metzger2022};
   (4) a failed/partial explosion of a WR star: a direct collapse of a WR star to a black hole launches a subrelativistic jet, and the jet interacts with a dense CSM releasing the radiated energy \citep[e.g.,][]{Perley2022}.
   \citet[][]{Gal-Yam2022} proposed a WR star to be the progenitor of SN~2019hgp based on its observational properties, and suggested a possibility that
   the differences between Type Ibn and Icn SNe arise from their different types of WR star progenitors: helium/nitrogen-rich WN stars for Type~Ibn SNe and C-rich WC stars for Type~Icn SNe. \citet[][]{Pellegrino2022} proposed that multiple progenitor channels could explain different Type~Icn SNe, based on the properties of the SNe and their explosion sites. They suggested that the progenitor of SN~2019jc was a low-mass, ultra-stripped star whereas those of SNe~2019hgp, 2021csp and 2021ckj were WR stars. Furthermore, \citet[][]{Davis2022} suggested a binary-stripped progenitor for SN~2022ann rather than a single massive WR progenitor.
  
   
   In this paper, we report photometric and spectroscopic observations of the Type Icn SN~2021ckj and discuss its observational properties as compared to those of SNe~2019hgp and 2021csp, the two most well-observed members of this class. SN~2021ckj was discovered as ZTF21aajbgol by the Zwicky Transient Facility \citep[ZTF;][]{Bellm2019} on 9.29 February 2021 UT (59254.29 MJD), which was reported by the Automatic Learning for the Rapid Classification of Events \citep[ALeRCE;][]{Forster2021}. The object was not detected down to a limiting magnitude of 20.7 mag on 7.36 February 2021 UT (59252.36 MJD), as constrained by the Asteroid Terrestrial impact Last Alert System \citep[ATLAS;][]{Tonry2018,Smith2020}. We adopt the middle point between the discovery and the last non-detection, 59253.38 MJD, as the explosion date. The phases in this paper are shown in rest-frame days with respect to this explosion date. The spectroscopic classification of SN~2021ckj as a Type Icn SN was conducted on 16 February 2021 by \citet[][]{Pastorello2021}, based on a spectrum taken with the ESO Very Large Telescope (VLT) and the FORS2 spectrograph.
   
   Throughout this paper, we adopt a redshift $z=0.141$ (measured from the narrow host-galaxy emission lines), and a distance modulus $\mu=39.04$ mag (assuming H$_{0}=73$ km s$^{-1}$ Mpc$^{-1}$, $\Omega_{m}=0.27$ and $\Omega_{\Lambda}=0.73$). Since the spectra of SN~2021ckj show no signs of narrow Na~I~D interstellar absorption, we assume that it has a minimal host-galaxy extinction (see Section~\ref{sec:spec}). Only the Galactic extinction correction is performed to the photometric and spectroscopic data, assuming $E(B-V)=0.049$ \citep[][]{Schlafly2011}, $R_V=3.1$ and the extinction curve by \citet[][]{Cardelli1989} using IRAF \citep[][]{Tody1986,Tody1993}.


\section{Observations} \label{sec:obs}

\subsection{Photometry} \label{sec:photo}

SN~2021ckj was observed by several wide-field transient surveys, e.g., ZTF, ATLAS and the Panoramic Survey Telescope and Rapid Response System \citep[Pan-STARRS;][]{Chambers2016}. We obtained $g$-, $r$- and $i$-band images taken by ZTF through the NASA/IPAC Infrared Science Archive\footnote{\url{https://irsa.ipac.caltech.edu/}}. We also used the `cyan'-band photometry reported by ATLAS, and `white'-band photometry by Pan-STARRS.
We conducted multi-band (\textit{uBVgRriz}) photometry of SN~2021ckj with the EFOSC2 instrument mounted on the New Technology Telescope (NTT) at La Silla Observatory in Chile as a part of the extended Public ESO Survey for Transient Objects \citep[ePESSTO+;][]{Smartt2015}, as well as the AFOSC instrument mounted on the Copernico Telescope. The observation log is shown in Table~\ref{obs_log}.

We reduced the data using the PESSTO pipeline \citep[][]{Smartt2015} and IRAF, performing standard tasks such as bias-subtraction and flat-fielding. 
For the EFOSC2 data, we performed point-spread function (PSF) photometry after host galaxy subtraction with reference images taken on 3rd February 2022, while for the other data we conducted PSF photometry with the stacked Pan-STARRS $g$-, $r$- and $i$-band images as reference images.
A correction for the Galactic extinction was applied. The resulting photometry is provided in Table~\ref{tab:photo2}.

   \begin{table*}
      \caption[]{Log of the photometry of SN~2021ckj.}
         \label{obs_log}
     $$
         \begin{tabular}{cccccc}
            \hline
            \noalign{\smallskip}
            Date (UT) & MJD & Phase & Bands & Telescope & Instrument\\
            \noalign{\smallskip}
            \hline
            \noalign{\smallskip}
            2021 February 06.34 & 59251.34 & -1.79 & \textit{i} & 48 inch Samuel Oschin Telescope & ZTF Observing System\\
            2021 February 07.29 & 59252.29 & -0.96 & \textit{gr} & 48 inch Samuel Oschin Telescope & ZTF Observing System\\
            2021 February 09.29 & 59254.29 & +0.80 & \textit{gri} & 48 inch Samuel Oschin Telescope & ZTF Observing System\\
            2021 February 11.31 & 59256.31 & +2.57 & \textit{gr} & 48 inch Samuel Oschin Telescope & ZTF Observing System\\
            2021 February 12.30 & 59257.30 & +3.44 & \textit{i} & 48 inch Samuel Oschin Telescope & ZTF Observing System\\
            2021 February 15.30 & 59260.30 & +6.06 & \textit{gri} & 48 inch Samuel Oschin Telescope & ZTF Observing System\\
            2021 February 17.94 & 59262.94 & +8.38 & \textit{uBVgriz} & 1.82m Copernico Telescope & AFOSC\\
            2021 February 18.23 & 59263.23 & +8.63 & \textit{gri} & 48 inch Samuel Oschin Telescope & ZTF Observing System\\
            2021 February 20.27 & 59265.27 & +10.42 & \textit{gr} & 48 inch Samuel Oschin Telescope & ZTF Observing System\\
            2021 February 21.19 & 59266.19 & +11.23 & \textit{i} & 48 inch Samuel Oschin Telescope & ZTF Observing System\\
            2021 February 22.15 & 59267.15 & +12.07 & \textit{V} & 3.58m NTT telescope & EFOSC\\
            2021 February 22.27 & 59267.27 & +12.17 & \textit{g} & 48 inch Samuel Oschin Telescope & ZTF Observing System\\
            2021 February 23.24 & 59268.24 & +13.02 & \textit{BVRi} & 3.58m NTT telescope & EFOSC\\
            2021 February 24.27 & 59269.27 & +13.93 & \textit{gr} & 48 inch Samuel Oschin Telescope & ZTF Observing System\\
            2021 February 25.19 & 59270.19 & +14.73 & \textit{i} & 48 inch Samuel Oschin Telescope & ZTF Observing System\\
            2021 February 26.27 & 59271.27 & +15.68 & \textit{r} & 48 inch Samuel Oschin Telescope & ZTF Observing System\\
            2021 March 6.15 & 59279.15 & +22.59 & \textit{BVRi} & 3.58m NTT telescope & EFOSC\\
            2021 March 13.13 & 59286.13 & +28.70 & \textit{BVRi} & 3.58m NTT telescope & EFOSC\\
            2021 March 23.09 & 59296.09 & +37.43 & \textit{BVRi} & 3.58m NTT telescope & EFOSC\\
            2022 January 3.31 & 59582.31 & +288.28 & \textit{BVRi} & 3.58m NTT telescope & EFOSC\\
            2022 February 3.25 & 59613.25 & +315.40 & \textit{BVRi} & 3.58m NTT telescope & EFOSC\\
            \noalign{\smallskip}
            \hline
        \end{tabular}
     $$
   \end{table*}

   \begin{table*}
      \caption[]{Photometry of SN~2021ckj}
         \label{tab:photo2}
     $$
         \tiny
         \begin{tabular}{ccccccccc}
            \hline
            \noalign{\smallskip}
            Phase & $u$ & $B$ & $V$ & $g$ & $R$ & $r$ & $i$ & $z$\\
            \noalign{\smallskip}
            \hline
            \noalign{\smallskip}
            -1.79 &-&-&-&-&-&-& > 20.375 &- \\
            -0.96 &-&-&-& > 21.641 &-& > 21.352 &-&- \\
            +0.80 &-&-&-& 19.647 (0.065) &-& 19.983 (0.091) & 19.994 (0.127) &- \\
            +2.57 &-&-&-& 19.256 (0.057) &-& 19.456 (0.088) &-&- \\
            +3.44 &-&-&-&-&-&-& >18.313 &- \\
            +6.06 &-&-&-& 19.508 (0.086) &-& 19.253 (0.136) & 19.682 (0.129) &- \\
            +8.38  & 20.551 (0.073) & 20.459 (0.126) & 19.939 (0.070) & 20.009 (0.073) &-& 19.966 (0.034) & 19.931 (0.036) & 19.981 (0.083)\\
            +8.63 &-&-&-& 20.164 (0.194) &-& 19.994 (0.308) & 20.129 (0.311) &- \\
            +10.42 &-&-&-& 20.659 (0.160) &-& 20.551 (0.105) &-&- \\
            +11.23 &-&-&-&-&-&-& >19.865 &- \\
            +12.07 &-&-& 20.554 (0.065) &-&-&-&-&-\\
            +12.17 &-&-&-& 20.990 (0.400) &-&-&-&- \\
            +13.02 &-& 21.451 (0.052) & 20.824 (0.028) &-& 20.674 (0.025) &-& 20.784 (0.022) &-\\
            +13.93 &-&-&-& >20.580 &-& >17.340 &-&- \\
            +14.73 &-&-&-&-&-&-& >20.390 &- \\
            +15.68 &-&-&-&-&-& >20.314 &-&- \\
            +22.59 &-& 22.235 (0.079) & 21.637 (0.069) &-& 21.558 (0.066) &-& 21.759 (0.094) &-\\
            +28.70 &-& 22.312 (0.061) & 21.757 (0.065) &-& 21.664 (0.063) &-& 21.871 (0.101) &-\\
            +37.43 &-& 22.476 (0.053) & 22.049 (0.040) &-& 22.081 (0.051) &-& 22.283 (0.076) &-\\
            +288.28 &-& >23.579 & >22.853 &-& >22.744 &-& >22.258 &-\\
            +315.40 &-& >23.691 & >22.837 &-& >22.765 &-& >22.278 &-\\
            \noalign{\smallskip}
            \hline\\
            \multicolumn{9}{l}{Notes. BVR-band photometry is calibrated to Vega magnitudes, while ugriz-band photometry to AB magnitudes.}
        \end{tabular}
     $$
   \end{table*}

\subsection{Spectroscopy} \label{sec:spec}

The first spectrum of SN~2021ckj was obtained by \citet[][]{Pastorello2021} at Phase +7.7 days with the FORS2 instrument \citep[][]{appenzeller98} at the ESO VLT with the 300V grism and a 1 arcsecond slit width.
We took another spectrum at Phase +12.1 days with EFOSC2/NTT using the Gr\#13 grism and a 1 arcsecond slit width as a part of ePESSTO+ collaboration \citep[][]{Smartt2015}. 
In addition, we obtained one more spectrum at Phase +21.2 d using FORS2/VLT with grism 300V.

The EFOSC2 and FORS2 data were reduced using the PESSTO\footnote{\url{https://github.com/svalenti/pessto}} and ESOReflex \citep[][]{Freudling2013} pipelines, respectively, which include standard tasks such as bias-subtraction, flat-fielding and a wavelength calibration based on arc frames.
The flux calibration was performed using observations of a spectrophotometric standard star.
%

\section{Results}  \label{sec:results}
 
\subsection{Photometric properties}  \label{sec:photo_prop}

   \begin{figure*}
   \centering
   \includegraphics[width=\hsize]{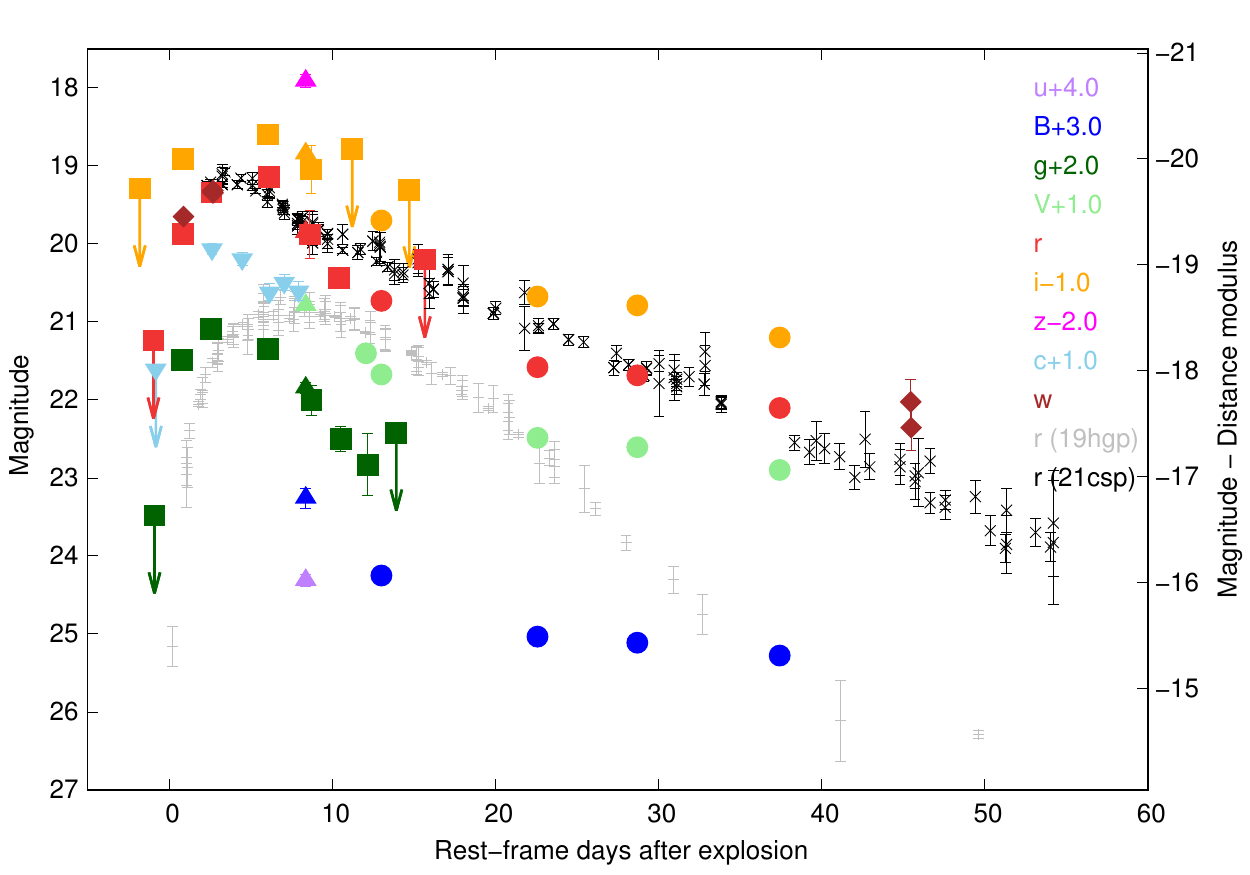}
      \caption{The optical light curves of SN~2021ckj. The data are taken with the 1.82m Copernico Telescope (triangles), NTT (circles), ZTF (filled squares), ATLAS (inverted triangles) and Pan-STARRS (diamonds). The $R$-band magnitudes in the NTT data are converted into r-band magnitudes using the relation in \citet{Chonis2008}.
      The gray and black points show the r-band light curves of SNe~2019hgp \citep[][]{Gal-Yam2022} and 2021csp \citep[][]{Fraser2021}, respectively, being plotted using their absolute magnitudes. Limiting magnitudes are indicated with arrows.
      %
      %
      All the magnitudes are corrected with the Galactic extinction \citep[$E(B-V)=$0.027 and 0.019 mag for SNe~2021csp and 2019hgp, respectively;][]{Schlafly2011}.
      The error bars for all data points are also plotted, even though they are smaller than the sizes of the symbols in most cases.
              }
      \label{fig:LC}
   \end{figure*}
   
Figure~\ref{fig:LC} shows multi-band LCs of SN~2021ckj. 
They show bright peak brightness and rapid rises and declines. The absolute peak magnitudes are $\sim -20$ mag in the optical bands (e.g., \textit{g}, cyan and \textit{r} bands). The rise time and the time above half-maximum ($t_{1/2}$) are $\sim 4$ and $\sim 10$ days, respectively, in the \textit{g}/cyan bands. 
These features are similar to other Type~Icn SNe \citep[SNe~2019hgp, 2021csp and 2019jc;][]{Gal-Yam2022, Perley2022, Fraser2021, Pellegrino2022}. Especially, the LCs of SN~2021ckj are almost replicas of SN~2021csp around the peaks (see their r-band LCs in Figure~\ref{fig:LC}), although they deviate from each other at later phases. 

   \begin{figure}
   \centering
   \includegraphics[width=\hsize]{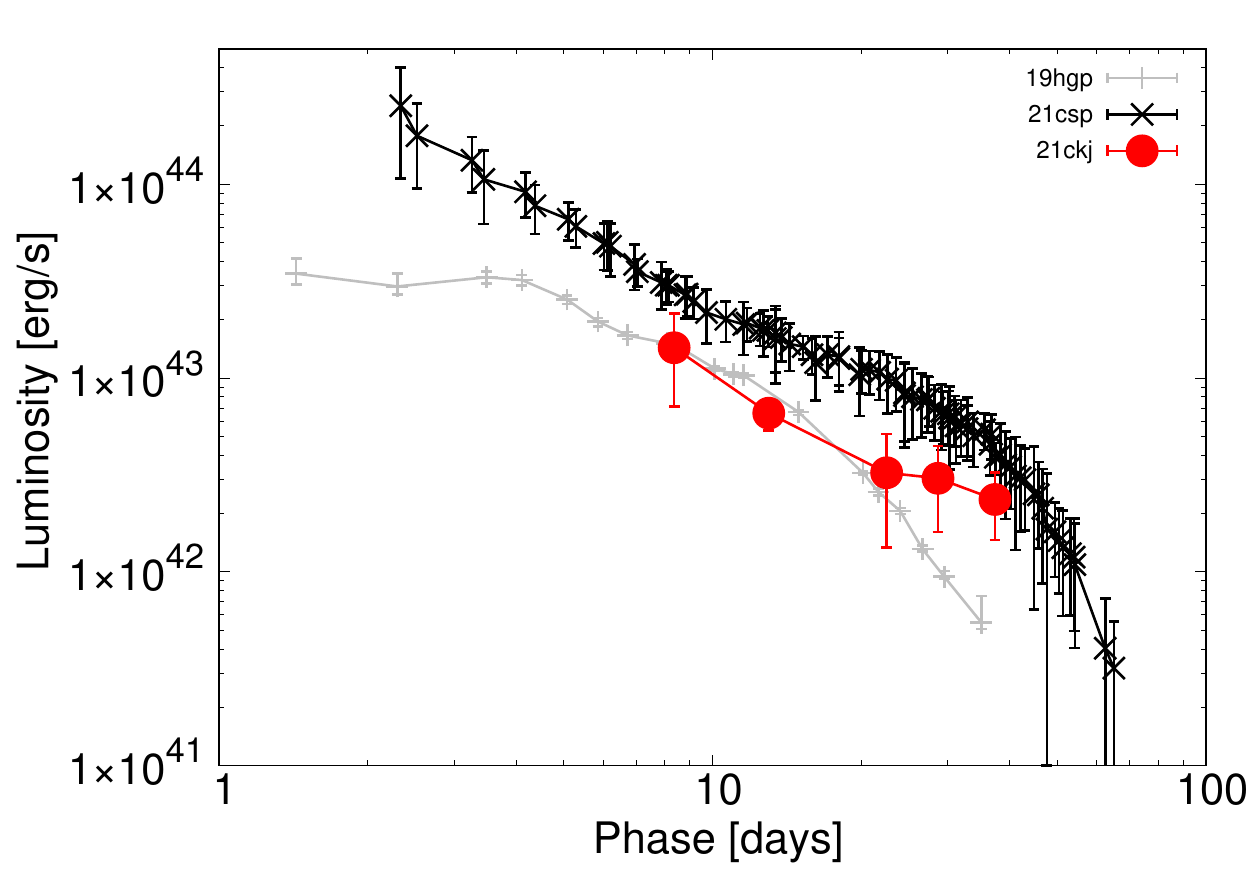}
   \includegraphics[width=\hsize]{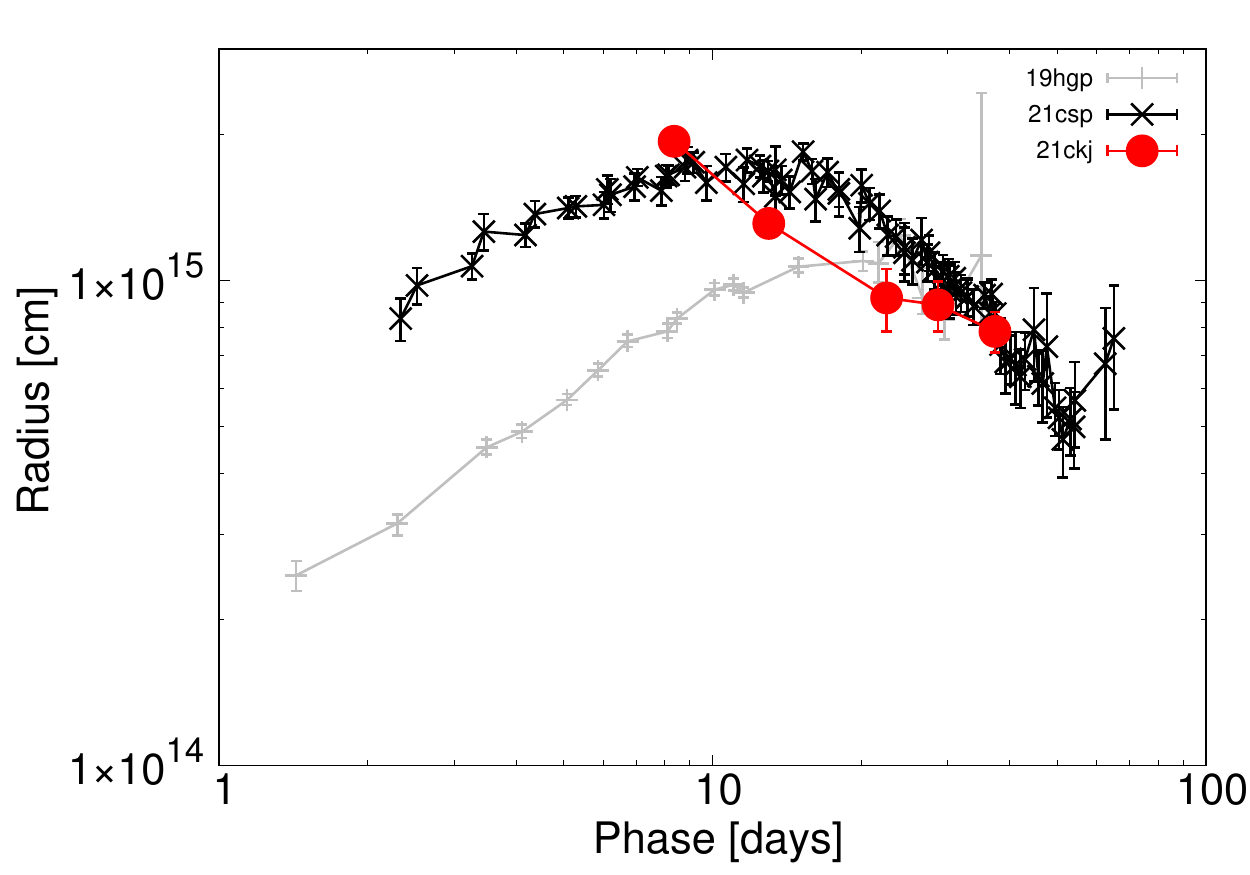}
   \includegraphics[width=\hsize]{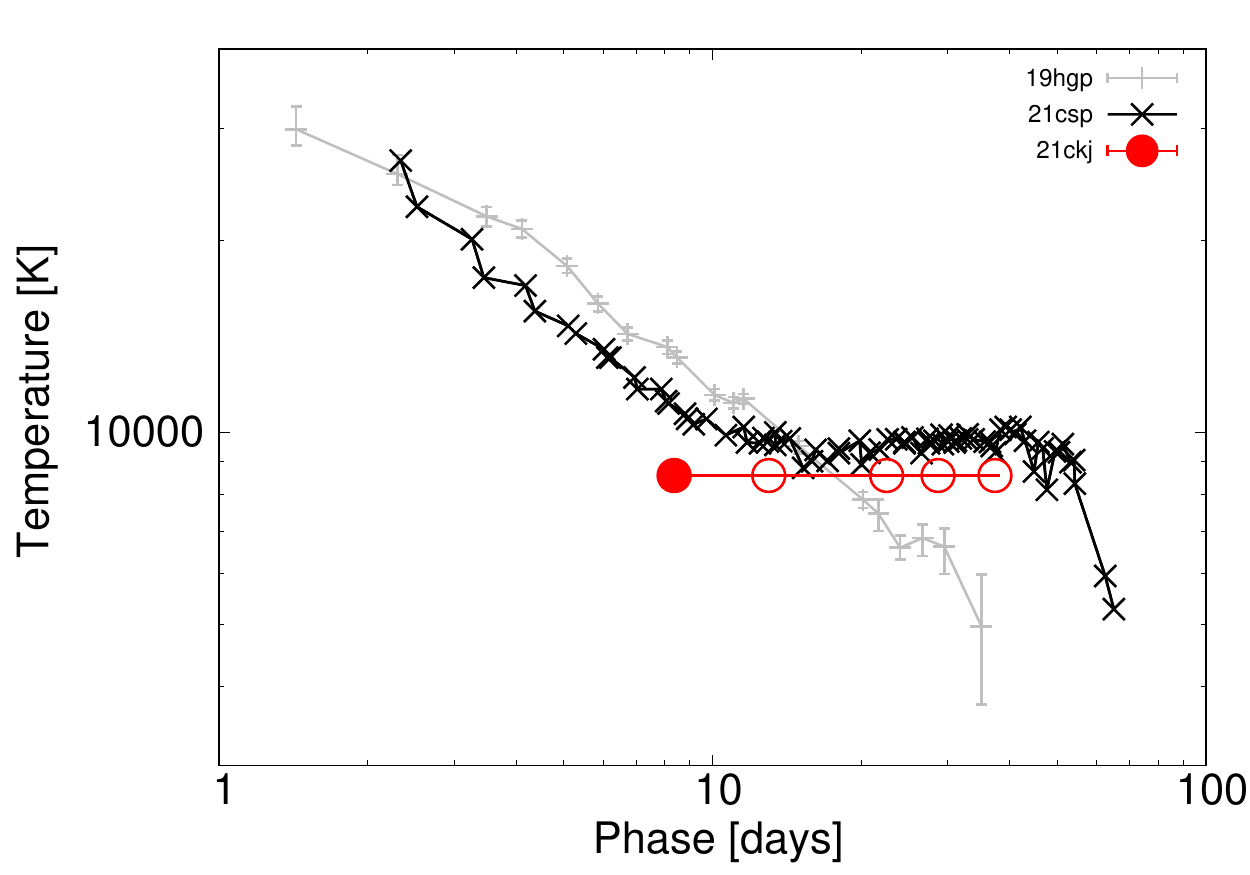}
      \caption{Evolution of the photospheric parameters estimated from blackbody fitting to the photometry of SN~2021ckj (red points connected with a line). Open red circles are the assumed values. Comparison objects, SNe~2019hgp and 2021csp, are shown with the gray and black points connected with lines, respectively. The values for the comparison objects were taken from \citet[][]{Gal-Yam2022,Fraser2021}. 
              }
         \label{fig:BB_param}
   \end{figure}

First, we calculated black-body (BB) radius and temperature for the SN radiation at Phase~+8.38 through $\chi^{2}$ fits of the $uBVgriz$-band photometry with the Planck function. Since the photometric data in the other phases are limited, the wavelength ranges of the observations do not cover the peaks of BB curves with anticipated temperatures ($\sim 10000$ K). In fact, similar $\chi^{2}$ fittings for the other phases produce poor constraints on their BB radii and temperatures. Therefore, we performed the $\chi^{2}$ fittings with the fixed temperature estimated from Phase~+8.38 for the other phases. Here, we did not use the data at Phase~12.07, which has only one band of the photometry.
From the derived BB parameters, we calculated the bolometric luminosity by integrating the BB function across wavelengths. The estimated values are shown in Figure~\ref{fig:BB_param}. The derived luminosities, radii, and temperatures for SN~2021ckj are very similar to those for SN~2021csp, while they have different evolution from those of SN~2019hgp.

\subsection{Spectroscopic properties} \label{sec:spec_prop}

   \begin{figure*}
   \centering
   \includegraphics[width=0.9\hsize]{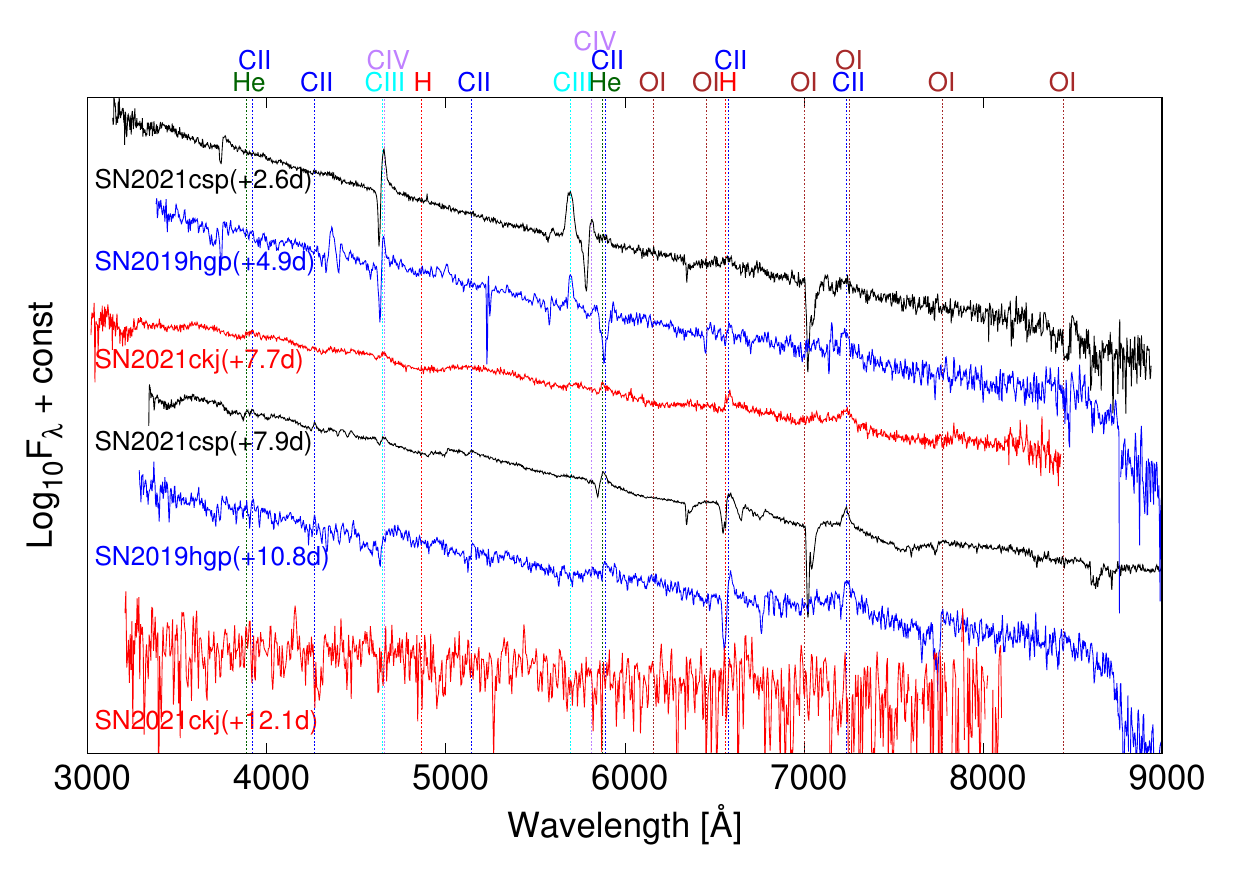}
   \includegraphics[width=0.9\hsize]{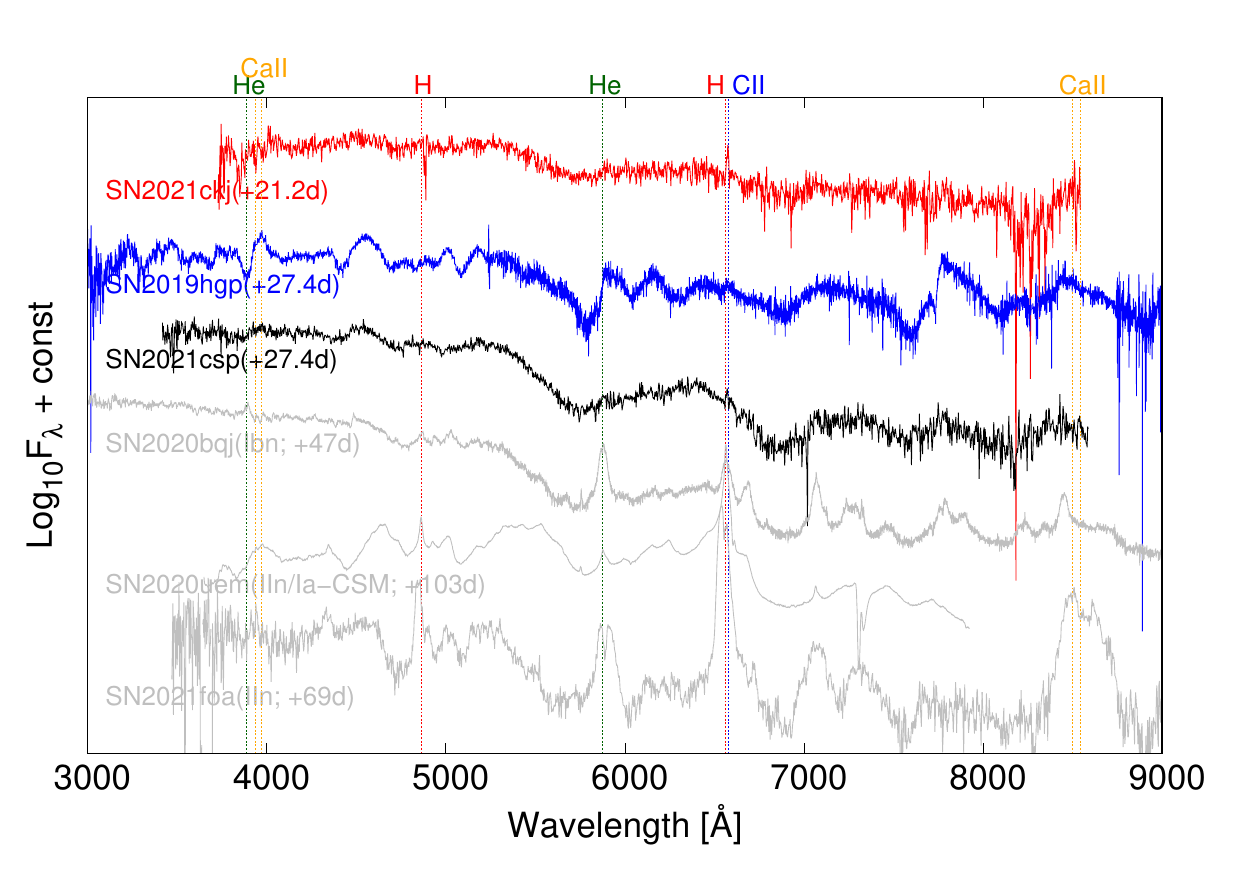}
      \caption{Top panel: the early spectra of SN~2021ckj (red), compared with the other Type~Icn SNe~2019hgp (gray) and 2021csp (black). Bottom panel: the same as the top panel but for the late-time spectra. For comparison, the spectra of SNe 2020bqj \citep[Ibn;][]{Kool2021}, 2020uem \citep[IIn/Ia-CSM;][]{Uno2023} and 2021foa \citep[IIn;][]{Reguitti2022} are plotted. These data were obtained through the Weizmann Interactive Supernova Data Repository \citep[WISeREP; \url{https://www.wiserep.org/};][]{Yaron2012}.}
         \label{fig:spec}
   \end{figure*}

In the early spectrum of SN~2021ckj (Phase +7.7 days), there are highly-ionized narrow lines of C~II/C~III, which are also seen in the spectra of SNe~2019hgp and 2021csp at similar epochs (Phases +10.8 and +7.9 days, respectively; see Figure~\ref{fig:spec}). This indicates that the CSM has a similar ionization state at similar epochs for these SNe. SN~2021ckj shows minimal absorption for strong C~II lines (e.g., C~II $\lambda$5890, 6578), while SNe~2019hgp and 2021csp show an evident blue-shifted absorption component for the C~II lines. 
The second spectrum of SN~2021ckj (Phase +12.1 days) is smooth and featureless similar to other Type~Icn SNe around this phase, although the signal-to-noise ratio of the spectrum is not high (see Figure~\ref{fig:spec}).

The late-time spectrum of SN~2021ckj (Phase +21.2 days) shows a blue pseudo-continuum due to Fe lines with some undulations at $\lambda \lesssim 5500$~{\AA} (so-called "Fe bump"), while no evident lines are visible in the red part except for Ca~II NIR triplet (see Figure~\ref{fig:spec}). This Fe bump is also seen in other interacting SNe, e.g., Type Ibn \citep[e.g.,][]{Pastorello2016}, Type~Ia-CSM \citep[e.g.,][]{Fox2015}, some Type~IIn SNe \citep[e.g.,][]{Turatto1993} and interacting Type-Ic SNe \citep[][]{Kuncarayakti2018,kuncarayakti2022}.
On the other hand, SNe~2019hgp and 2021csp show more line features while showing a similar "Fe bump". The narrow line seen near the H$\alpha$ wavelength in the late-time spectrum of SN~2021ckj could be identified with C II $\lambda$6578.

It is intriguing that the overall spectra are similar among these Type~Icn SNe while the difference is seen in the line profiles and velocities, both in the early and late phases. This is one of the topics that we address in the present work; we suggest that the difference seen in the line features might be due to different ejecta velocities and/or viewing angle effects due to an aspherical explosion.

\section{LC modeling} \label{sec:LC_modeling}
In this section, we model the bolometric LCs of SNe~2021ckj, 2021csp and 2019hgp using the CSM interaction model of \citet[][]{Maeda2022}, in order to estimate their ejecta and CSM properties. In the calculations, we adopt the following broken power law for the density structure of the SN ejecta as a function of velocity ($v$); $\rho_{\rm{SN}} \propto v^{-n}$ and $n=7$ for the outer part, and a constant density for the inner part. The normalization in the density is set by specifying the ejecta mass ($M_{\rm{ej}}$) and the explosion energy ($E_{K}$). For the CSM, a single power-law function of the distance ($r$) is assumed using constant values, $D$, $D^{'}$ and $s$;  $\rho_{\rm{CSM}} = D r^{-s} = 10^{-14} D^{\prime} (r/5\times10^{14} \; \rm{cm})^{-s} \; \rm{g}\;\rm{cm}^{-3}$. For Type Ibn SNe, \citet[][]{Maeda2022} derived $s \sim 2.5 - 3.0$ and $D^{\prime} \sim 0.5 -5$. The mass-loss rate responsible for the CSM at the reference radius, $5 \times 10^{14}$ cm (which corresponds to the position of the shock wave on day 2 for the shock velocity of $30,000$ km s$^{-1}$), is $\dot M \sim 0.05 D^{\prime} (v_{\rm w} / 1,000 \ {\rm km} \ {\rm s}^{-1}) M_\odot$ yr$^{-1}$ where $v_{\rm w}$ is the mass-loss wind velocity. We assume a C+O-rich composition, but the optical LC models explored here are not sensitive to the choice of the composition as the main power in the model is provided by the free-free emission at the high-temperature forward shock. Also, we note that our models are not aimed to provide a unique solution for the model parameters. 

The model LCs and characteristic-radius evolution are shown in Figure~\ref{fig:LC_model}. The radius shown here is the position of the contact discontinuity, i.e., representative scale of the interacting region, which provides the maximum radius expected in the interaction model. The photosphere radius is expected to be either close to this radius (if the shocked region is optically thick to the optical photons) or smaller (if the shocked region is optically thin). Note that the model assumes spherical symmetry both in the ejecta and the CSM. Possible effects of asymmetry in the ejecta, as indicated by the spectral features in these objects, will be discussed in Section~\ref{sec:discussions}.

\subsection{SN~2019hgp}
Our first attempt is made for SN 2019hgp. Given the close similarity of its LC to the Type~Ibn SN template (see Figure~\ref{fig:LC_model}), a good match to its LC is obtained with the ejecta and CSM properties similar to those applied for Type~Ibn SNe by \citet[][$M_{\rm{ej}} \sim 2-6$ M$_\odot$, $E_{K}\sim 1 \times 10^{51}$ ergs, $s \sim 2.5-3.0$, and $D^{\prime} \sim 0.5-5.0$]{Maeda2022}; $M_{\rm{ej}} = 3 M_\odot$, $E_{K} = 2.5 \times 10^{51}$ ergs, $s=2.9$, and $D^{\prime} = 2.3$ for SN~2019hgp (blue line). The evolution of the radius of the shocked region roughly follows the BB radius up to $\sim 10-20$ days (noting that we do not model in detail the first 4 or 5 days, as this will require a more detailed treatment of radiation transfer effects). Interestingly, the photosphere starts receding at $\sim 10 - 20$ days, which coincides with the kink seen in the LC evolution. This transition in the LC from a flat to steep evolution is interpreted to be caused by the change in the forward-shock property from the optically-thick cooling phase to the optically-thin adiabatic phase \citep[see][]{Maeda2022}. According to this interpretation, we expect that the photosphere is formed in the shocked region in the earlier phases but it later recedes into the ejecta once the shock becomes optically thin. This interpretation is consistent with the spectral evolution from the CSM interaction-dominated phase (characterized by a featureless blue continuum with narrow lines) into the SN ejecta-dominated one (with a broad-line spectrum; see Figure~\ref{fig:spec}).
The simultaneous occurrence of the accelerated LC decay and the spectral evolution due to the receding photosphere, is predicted by the model of \citet[][]{Maeda2022}, strengthening the case that SN 2019hgp is mainly powered by the SN-CSM interaction.

\subsection{SNe~2021csp and 2021ckj}
The LC of SN 2021csp shows an initial rapid-decay phase, followed by a flattening ($> 10$ days) and then again by a steep decay ($> 40$ days). SN 2021ckj might show similar evolution to SN 2021csp, even though the data are not well enough sampled in the earliest and latest phases. Since the photometric and spectroscopic properties of SNe 2021csp and 2021ckj are very similar (see Section~\ref{sec:results}), 
we assume that they are "twins" and model the LC of SN~2021csp for understanding both SNe.
The initial decay is not predicted directly in the SN-CSM interaction model that assumes a single power-law CSM distribution. Thus, we discuss this part separately in Section~\ref{sec:4.3}.
The LC evolution, except for this initial decay ($\gtrsim 10$ days), is similar to the evolution seen in the Type~Ibn SN template, but being somewhat brighter and with a slower transition. The model shown in Figure~\ref{fig:LC_model} (red line) has the following parameters; $M_{\rm{ej}} = 4 M_\odot$, $E_{K} = 4 \times 10^{51}$ ergs, $s=2.9$, and $D^{\prime} = 5.1$. 
%
%
As compared to the models for Type Ibn SNe and SN~2019hgp, the ejecta and CSM properties of these SNe are slightly different: the energy per ejecta mass is larger and the CSM density is on a higher side, although the total ejecta mass and the CSM density distribution are similar.


\subsection{The initial rapid-declining phase of SN~2021csp} \label{sec:4.3}
As mentioned above, in the early phase, the LC of SN~2021csp shows different properties from those of SN~2019hgp and Type~Ibn SNe. It exhibits a rapidly declining light curve after maximum as well as a very high photospheric velocity of $\sim 30,000$ km s$^{-1}$, estimated from the time evolution of the black-body radius (cyan line in the right-hand panel of Figure~\ref{fig:LC_model}). In addition, the photosphere starts receding already at $\sim 5$ days, which is earlier than seen for SN~2019hgp ($\sim 20$ days; see Figure~\ref{fig:LC_model}). These properties are difficult to reconcile with those at later phases in the present CSM-interaction model. In particular, the initial very high photospheric velocity contradicts the bright and slow LC evolution at later phases. Although the latter requires a large amount of CSM, this should substantially decelerate the shock wave, causing a challenge in reproducing the early high velocity. This may indicate that the initial phase is powered by a different mechanism. We consider two scenarios; (1) a CSM interaction by a highly energetic ejecta component in addition to the interaction by slower ejecta at later phases, and (2) the initial phase is not powered by instantaneous interaction, but by a mechanism similar to the shock-cooling emission. In the former scenario, it is assumed that the SN-CSM interaction keeps ongoing in optically thin environment, and the kinetic energy dissipated at the shock front is immediately converted to optical radiation as it is observed. In the latter scenario, most of the kinetic energy is dissipated early on before the observation, either within the progenitor's envelope or optically-thick CSM, after that the dissipated energy is converted to thermal energy and forms an expanding fireball; the subsequent radiation loss  in the cooling fireball will produce the characteristic `shock-cooling' emission.
%

\subsubsection{CSM interaction scenario}
To support scenario (1), we show an additional model (magenta line) in which only $E_{K}$ is changed from the reference model for the later phase of SNe~2021csp and 2021ckj, while the other parameters are unchanged (i.e., adopting different ejecta properties but the same CSM properties). The energy, $E_{K}$, is set to allow the radius of the shocked region to expand with $v\sim 30,000$ km s$^{-1}$. This is realized if $E_{K} \sim 35 \times 10^{51}$ erg (for fixed ejecta mass of $4 M_\odot$). This model can roughly explain the rapid initial decay. In this energetic model, the shocked region is in the optically-thin adiabatic regime \citep[][]{Maeda2022}. Qualitatively, the situation considered here is the following: an asymmetric and highly-energetic ejecta component is ejected toward a specific direction within a limited solid angle (magenta), which is followed by the nearly spherical slower ejecta component (red). These two components interact with nearly spherically-distributed CSM. The combination of these two components can qualitatively explain the LC behavior. Interestingly, the energy is similar to those derived for SNe associated with gamma-ray bursts \citep[e.g.,][]{Cano2011}; this may not be surprising due to the initially high-velocity photosphere seen in SNe 2021csp and 2021ckj. We note that the energy and the mass in the model are probably overestimated, if we consider a collimated outflow. There are two caveats in this scenario. First, the initial phase is a bit too bright as compared to the data, while this may simply be modified by including a more realistic treatment of the ejecta geometry. Second, probably more importantly, this model predicts that the initial decay is in the optically-thin adiabatic phase of the energetic component, and the straightforward expectation is that the photosphere already starts receding from the beginning. This might be inconsistent with the initial rapid expansion of the photosphere, while investigation of further details will require more sophisticated treatment of radiation transfer effects.

\subsubsection{Shock-cooling scenario}
\begin{figure*}
   \centering
   \includegraphics[width=0.49\hsize]{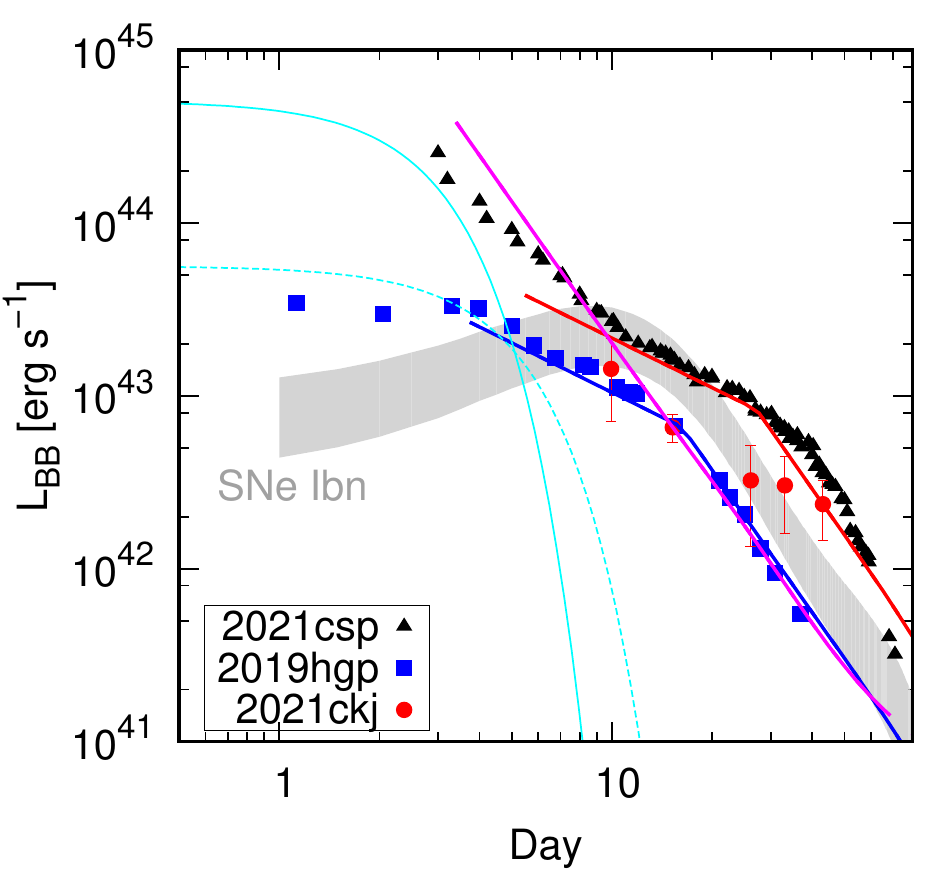}
   \includegraphics[width=0.49\hsize]{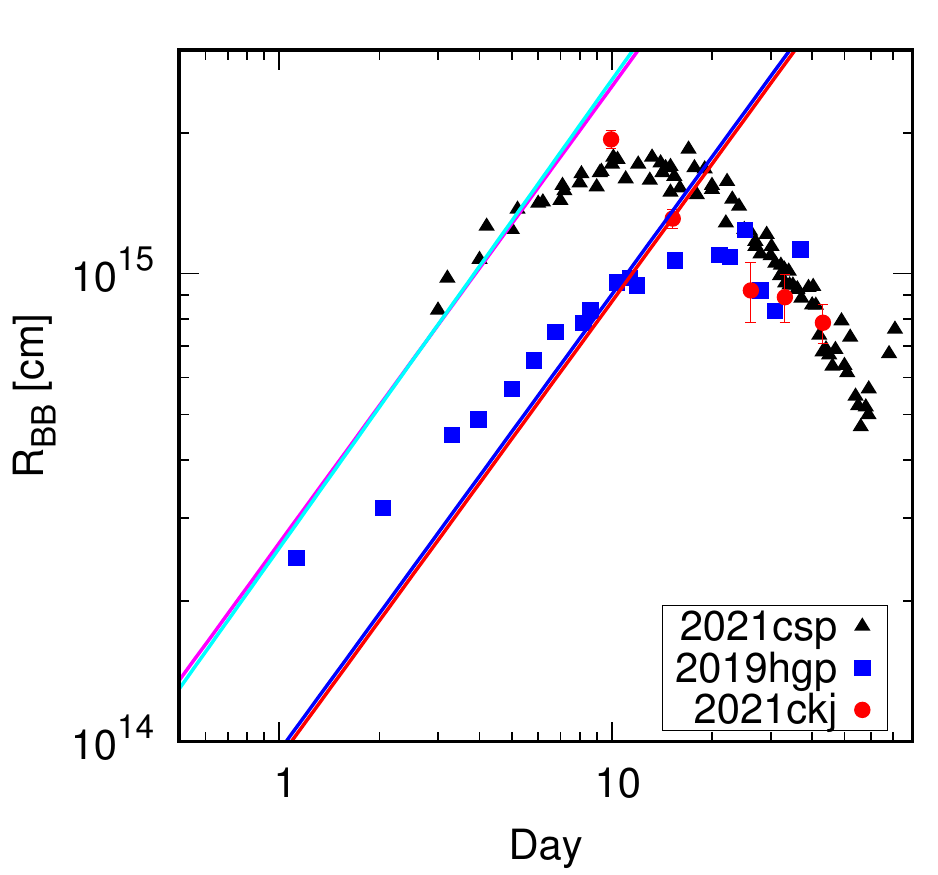}  
      \caption{Light curve models (left panel) and evolution of the radius at the contact discontinuity, as compared to the photospheric radius (right panel). The model parameters are as follows; $M_{\rm{ej}} = 3 M_\odot$, $E_{K} = 2.5 \times 10^{51}$ ergs, $s=2.9$, $D^{\prime} = 2.3$ for SN 2019hgp (blue lines), and $M_{\rm{ej}} = 4 M_\odot$, $E_{K} = 4 \times 10^{51}$ ergs, $s=2.9$, $D^{\prime} = 5.1$ for SNe 2021ckj and 2021csp (red lines). Additionally, two scenarios for the initial decay seen in SN 2021csp are shown; (1) the SN-CSM interaction driven by a highly-energetic component (magenta lines) and (2) the shock-cooling of an energetic component (solid-cyan lines). As a comparison, the same but for a canonical explosion energy is also shown (dashed-cyan line). The gray hatching shows the template of the light curves of Type~Ibn SNe \citep[][]{Maeda2022,Hosseinzadeh2017}.
              }
         \label{fig:LC_model}
   \end{figure*}

Alternatively, the shock-cooling scenario may provide a more natural explanation. From the reference model (red line) in Figure~\ref{fig:LC_model}, we see that the radiation starts diffusing out at 5 days at $\sim 5 \times 10^{14}$ cm, signaling the `CSM breakout'. The model is not sensitively affected by the CSM distribution below this radius. Indeed, if the steep CSM density distribution is truncated toward the inner region generating an additional dense CSM (or envelope) component in the innermost region (i.e., the high-density component surrounded by a relatively flat CSM, which is then followed by the steep decrease starting at $\sim 5 \times 10^{14}$ cm), it would create `shock-cooling' emission. 

We consider the simplified (CSM) shock-cooling model of \citet[][]{Maeda2018}, which follows the formalism by \citet[][]{Arnett1980, Arnett1982}. The parameters are the CSM mass and radius ($M_{\rm  CSM}$ and $R_{\rm CSM}$) for the confined CSM (or the extended envelope), and $V_{\rm sh}$, which is the shocked-shell velocity after the breakout. Accordingly, the energy dissipated by this interaction (i.e., the ejecta kinetic energy above $V_{\rm sh}$) is approximated by $E (> V_{\rm sh}) \sim 0.5 M_{\rm CSM} V_{\rm sh}^2$. For demonstration purposes, we fix the opacity to be 0.1 g cm$^{-3}$ and the thickness of the shocked region to be equal to the shock radius. In this model, we expect that the photospheric radius follows the expansion of the shell with $V_{\rm sh}$, and thus we set $V_{\rm sh} = 30,000$ km s$^{-1}$. The model in Figure~\ref{fig:LC_model} (solid-cyan) is obtained with $R_{\rm CSM} = 10^{13}$ cm and $M_{\rm CSM} = 0.5 M_\odot$, hence $E (> 30,000 \ {\rm km} \ {\rm s}^{-1}) \sim 5 \times 10^{51}$ erg, which provides a rough representation of the initial rapid decay. This model requires a highly energetic shock, and this is not consistent with simple interpolation of the ejecta parameters adopted in the CSM-interaction model for the later phase. Therefore, we need a collimated high-energy ejecta component, which also decreases the necessary energy and mass budgets. 
For comparison, if we set $V_{\rm sh} = 10,000$ km s$^{-1}$, with $E (> 30,000 \ {\rm km} \ {\rm s}^{-1}) \sim 0.5 \times 10^{51}$ erg, to mimic the outermost region of the reference model but fixing the other parameters (e.g., the CSM properties), it would become much fainter (dashed-cyan). Interestingly, this cooling model roughly matches the initial phase of SN 2019hgp, indicating that such a confined CSM/envelope component could also exist behind SN 2019hgp. 

\subsection{Conclusions of the LC modeling}
From the above considerations, the following configuration may explain the LC evolution and the velocity evolution of SN 2021csp (and SN 2021ckj). The CSM distribution can be largely spherical, with the inner dense component within $\sim 10^{13}$ cm (which might be more like an envelope). The ejecta may be highly aspherical, composed of two components; a collimated high-energy outflow and a spherical canonical SN component. Both components first create the shock-cooling emission, which is dominated by the high-energy component. Once the shock-cooling emission quickly decays, the emission will be dominated by the SN-CSM interaction from the canonical (and spherical) component, since the high-energy component decays quickly also in the SN-CSM interaction. The photosphere is expected to first follow $V_{\rm sh}$, and once it becomes optically thin and the emission is dominated by the underlying SN-CSM interaction of the canonical and spherical SN component, the photosphere will eventually start receding toward the inner ejecta. 

On the one hand, the combination of the LC and velocity evolution of SN 2019hgp is consistent with a spherical SN-CSM interaction model, and there is no hint of an aspherical and high-energy ejecta component. On the other hand, as discussed above, the data for SNe 2021ckj and 2021csp indicate the existence of aspherical (potentially collimated) high energy component in either case: the CSM-interaction and shock-cooling models for the early phase.

\section{Spectral modeling} \label{sec:spec_modeling}

We use the modular open-source Monte Carlo radiative transfer code \texttt{TARDIS} \citep{Kerzendorf2014, Kerzendorf2022} to generate synthetic spectra for SN~2021ckj.  \texttt{TARDIS} is a one-dimensional code which for a user-defined ejecta composition and ejecta profile generates a time-independent synthetic spectrum. The code assumes that the ejecta are spherically symmetric, in homologous expansion and have an optically thick photosphere which emits r-packets (photon bundles) with energies sampled from a blackbody \citep{Kerzendorf2014}. We use similar techniques to \cite{Gillanders2020}, who used \texttt{TARDIS} to model the fast blue optical transient (FBOT) AT2018kzr. To produce a model to the observed spectra, we explore parameters controlling the ejecta composition, density profile, photospheric luminosity, and the explosion epoch to empirically obtain models that match the observations. We adopt a density profile $\rho (V)$ that is a power law in the ejecta speed ($V$). This power-law is assumed to extend from the inner boundary of the simulation domain ($V_{min}$) to the outer boundary ($V_{max}$), and is described by $\rho_{0}$ (which is a reference density), $t_{exp}$ (the adopted explosion epoch) and a power-law index $\Gamma$ via:
\begin{equation}
	\rho (V, t_{exp}) = \rho_{0} \left( \frac{t_{0}}{t_{exp}} \right)^3 \left ( \frac{V}{V_0}\right ) ^{-\Gamma},
\end{equation}
where we adopt reference constants $t_0 = 10$~days and $V_0 = 10,000$~km~s$^{-1}$.

We attempt to produce a self-consistent model with a uniform one-zone composition that reproduces the observed spectra of SN 2021ckj such that $\rm V_{ \rm min}$, $t_{exp}$ and the photospheric luminosity ($L_{phot}$) are the only parameters to change between the model spectra. To evolve a model forward in time, $t_{exp}$ increases and $\rm V_{ \rm min}$ is expected to decrease, corresponding to a recession of the photosphere into the inner ejecta. 
%
%

\begin{figure*}[!h]
    \centering
    \includegraphics[width=\hsize]{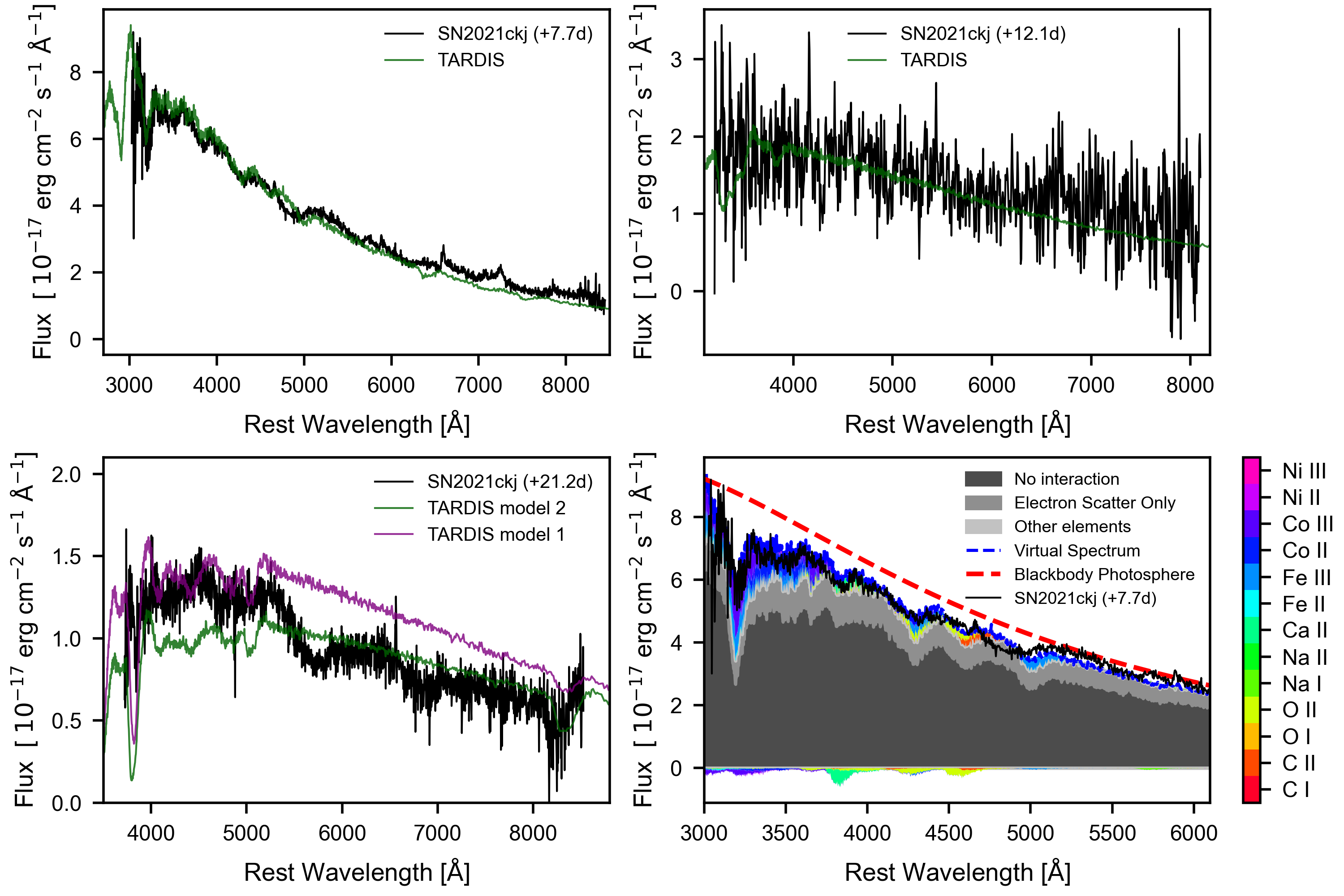}
    \caption{Top left panel: comparison of our best model (green) to the observed +7.7 day spectrum (black). Top right panel: Model spectrum (green) comparison to the +12.1 day spectrum (black). Bottom left panel: TARDIS models (green for model 1 and purple for model 2) compared to the +21.2 day spectrum (black). Bottom right panel: a +7.7 day spectrum Spectral element DEComposition (SDEC) plot showing the contributions of each chemical species to the synthetic spectrum. }
    \label{fig:TARDIS_MODELS}
\end{figure*}

The first spectrum of SN2021ckj (at +7.7 day) reveals a hot blue continuum, narrow features which are plausibly from CSM interaction (redwards of 6000 \AA) and broad absorption features in the blue. Our \texttt{TARDIS} modelling efforts for this epoch were focused on reproducing the features in the blue since these put the strongest constraints on the velocity of the ejecta and are not obviously affected by the signatures of the CSM interaction. The spectrum shows five prominent features with absorption minima at $ \sim 3200$, $ 3500 $, $3800 $, $ 4300 $, and $4900$ $ \rm  \AA$. 
%
%
We find that the feature at $ \sim 3800$ $ \rm \AA$ is Ca H\&K and the other features are combinations of Co III, Co II and Fe III. The identification of these species is shown in the Spectral element DEComposition (SDEC) plot in Figure \ref{fig:TARDIS_MODELS}, and the model parameters are listed in Tables\,\ref{tab:TARDIS_den} and \ref{tab:TARDIS_comp}. 
To model this  epoch we require t$_{ \rm exp }= +15.0\pm1.0$\,days, but our estimate for the time of the explosion from the light curve implies this spectrum should be at a phase of t$_{\rm exp }=+7.7$\,days. The velocity of $V_{ \rm min}=10,500$ km s$^{-1}$ is well constrained by the Co III, Fe III and Ca H\&K lines which forces our t$_{ \rm exp }$ to be beyond the ATLAS explosion constraint. This discrepancy suggests that the \texttt{TARDIS} model may not be capturing the full physical picture of the expanding ejecta. The SN ejecta could be aspherical or may not have been in homologous expansion throughout -- i.e. the need to adopt a high value for t$_{\rm exp }$ may be suggestive that the ejecta were initially faster but have decelerated to $10,500$ km s$^{-1}$ by this epoch. The output photospheric temperature of the \texttt{TARDIS} simulation is $\sim 12,500$ K, which is comparable to the +2.8 days spectrum of the FBOT AT 2018kzr \citep{Gillanders2020}. A reasonable model fit to the lines which appear to be formed in the expanding photosphere is achieved with the composition listed in Table\,\ref{tab:TARDIS_comp}.

The second spectrum (Phase +12.1 days) is noisy and cannot be used to place  constraints on the model composition or velocity. We include this spectrum in our modelling just to constrain the model temperature ($\sim 8500$ K) at this epoch.
 

The final spectrum of SN 2021ckj at +21.2 days shows P-Cygni features at $\sim$ 4000 $ \rm \AA$ and $\sim$ 8600 $ \rm \AA$ which are produced by the Ca H\&K lines and the Ca II near-infrared triplet, respectively. The flux peaks in the blue, and we suspect it is dominated by a pseudocontinuum flux due to a forest of Fe lines. In other Type Icn SNe the presence of this ``Fe bump" has been taken as evidence of strong CSM interaction and is thought to be driven by winds or shocked gas \citep{Perley2022}. \texttt{TARDIS} cannot treat these interaction effects and we calculate two models at this epoch, \texttt{model 1} with a boosted $L_{phot}$ to reproduce the Fe features in the Fe bump and \texttt{model 2} to fit the SED for wavelengths $\lambda >$5500 $ \rm \AA$. We measure the photospheric temperature ($\sim 6500$ K) from the \texttt{model 2} fit to the SED.

We find our model density profile is not consistent among all epochs. To produce reasonable agreement with the observed spectra, we must modify the density profile parameter $\rho_{0}$. We use  $\rho_{0}= 3.5 \times 10^{-12}$\,g\,cm$^{-3}$ for the early (+7.7-day) spectrum and $2.0 \times 10^{-13}$ g cm$^{-3}$ for the +12.1-day and +21.2-day spectra. The spectra of SN 2021ckj show signs of CSM interaction and the light-curve modelling in Section \ref{sec:LC_modeling} suggests the ejecta may be aspherical. \texttt{TARDIS} cannot treat CSM interaction or aspherical ejecta, so we acknowledge the limitations of our models to fully represent the ejecta of SN 2021ckj. The fact that we find an inconsistency in the density profile parameter $\rho_{0}$ also suggests that single zone, homologously expanding, spherical ejecta are not what we are observing. Therefore, \texttt{TARDIS} can only provide line identifications of the broad spectral features, approximate composition, and constraints on the ejecta velocity. We find that our the photospheric velocity of 10,000 km s$^{-1}$ is consistent with SN 2021csp and the broader population of typical Type Ic SNe, and that the composition is likely dominated by oxygen and carbon, while the iron group elements produce the strong absorption lines in the blue.

\begin{table*}[!h]
\caption{Density profile parameters and velocity ranges in the \texttt{TARDIS} models of SN 2021ckj. The density profile parameters $\rho_0$ and $\Gamma$ are not held constant but are consistent between second and third epochs. The photospheric velocity $V_{\rm min}$ is reduced between model spectra.}
\vspace{1em}
\centering	
\begin{tabular}{lllllllll}
\hline
\hline
\multirow{2}{*}{Phase  (days)} & \multirow{2}{*}{$t_{exp}$ (days)} & \multirow{2}{*}{$V_{\rm min}$ (km s$^{-1}$)} & \multirow{2}{*}{$V_{\rm max}$ (km s$^{-1}$)} & \multicolumn{4}{l}{ \hspace{1.2cm}       Density Profile Parameters} \\                       &                      &                     &                       & $V_{0}$ (km s$^{-1})$         & $t_{0}$ (days)      & $\rho_0$ (g cm$^{-3}$)    & $\Gamma$      \\

\hline

+7.7                    & +15                           & 10,500                  & 15,000                  & 10,000       & 10      & $3.5 \times 10^{-12}$       & 10         \\
+12.1                    & +19.4                         & 10,000                   & 15,000                  & 10,000       & 10      & $2 \times 10^{-13}$       & 10          \\
+21.2                   & +30.6                         & 9,500                   & 15,000                  & 10,000       & 10      & $2 \times 10^{-13}$       & 10         
\end{tabular}
\label{tab:TARDIS_den}
\end{table*}

\begin{table}[!h]
\caption{Relative mass fractions for the SN 2021ckj ejecta as constrained by the +7.7 day spectrum, the relative mass fraction of each element is held constant for all \texttt{TARDIS} models. }
\vspace{1em}
\centering	
\begin{tabular}{lc}
\hline
\hline
Element & Relative mass fraction \\
\hline
C       & 0.001         \\
O       & $\sim$ 0.8                    \\
Na      & 0.1               \\
Ca      & $1\times 10^{-3}$                  \\
Ti      & $1\times 10^{-4}$            \\
Fe      & 0.007                   \\
Co      & 0.0174                  \\
Ni      & 0.00174  \\ 
\hline 
\end{tabular}
\label{tab:TARDIS_comp}
\end{table}

\section{Discussions} \label{sec:discussions}
In this section, we discuss the ejecta and CSM properties of SN~2021ckj compared to the well-observed Type~Icn SNe 2019hgp and 2021csp. 
From the similarity of the photometric properties between SNe~2021ckj and 2021csp, the properties of their ejecta and CSM should be very similar. The LC modeling suggests that SNe~2021ckj and 2021csp have two ejecta components (an aspherical high-energy component and a spherical canonical component) and a relatively spherical CSM. The spherical ejecta and the CSM components with the following parameters can explain their late-time LC evolution: $M_{\rm{ej}} = 4 M_\odot$, $E_{K} = 4 \times 10^{51}$ ergs, $s=2.9$, and $D^{\prime} = 5.1$. Here, $s$ and $D^{\prime}$ are constants describing a single power-law density profile of the CSM according to $\rho_{\rm{CSM}} = D r^{-s}$. In addition to these `canonical' components, a highly-energetic component is required to explain the rapid initial decay and the photospheric expansion in the first 10 days. 

On the other hand, the photometric evolution of SN~2019hgp can be explained by a spherical SN-CSM interaction model alone. There is no sign of an aspherical and energetic ejecta component. We estimate the values for the ejecta and the CSM as follows: $M_{\rm{ej}} = 3 M_\odot$, $E_{K} = 2.5 \times 10^{51}$ ergs, $s=2.9$, and $D^{\prime} = 2.3$. 
The ejecta properties for SN~2019hgp are similar to those for Type~Ibn SNe \citep[$M_{\rm{ej}} \sim 2-6 M_\odot$, $E_{K} \sim 1.0 \times 10^{51}$ ergs; see][]{Maeda2022}. 

The canonical (spherical) ejecta component of SNe~2021ckj and 2021csp are more energetic than that of SN~2019hgp. This might be related to the interpretation that SN~2021ckj (and SN~2021csp) has the aspherical energetic component in addition to the canonical one.
The density distribution of the CSM is a common feature in these three SNe, and is consistent with those estimated for Type~Ibn SNe \citep[$s \sim 2.5-3.0$;][]{Maeda2022}. This might imply that the mass-loss mechanism is similar in Type~Icn (and also Ibn) SNe. The CSM mass of SN~2021ckj (and SN~2021csp; $D^{\prime} = 5.1$) is higher than that for SN~2019hgp ($D^{\prime} = 2.3$), even though these values are consistent with those of Type~Ibn SNe \citep[$D^{\prime} \sim 2.5-5.0$;][]{Maeda2022}.

Although SNe~2021ckj and 2021csp show almost the same behavior in their photometric evolution (see \ref{sec:photo_prop}), their spectral features are slightly different. As we discussed in Section \ref{sec:spec_prop}, the early spectrum of SN~2021ckj shows a similar ionization state of the ejecta as SN~2021csp, with a different absorption-to-emission ratio for C~II lines. This implies that some aspherical structures in the emitting regions, and the viewing angles for these SNe are different. This is consistent with the conclusion inferred from the LC modeling. 
We might have been looking at SN~2021ckj from the polar direction of the aspherical high-energy ejecta component. Since the CSM is quickly swept up to a larger distance in the polar direction than in the other directions, we have less amount of CSM along the line of sight. On the other hand, the viewing angle for SN~2021csp might have been relatively off-axis, where there is a larger amount of CSM above the interaction shock due to the slower propagation of the CSM interaction shock, creating the stronger absorption parts.

The difference in the late-phase spectra of SNe~2021ckj and 2021csp might also support this scenario. Their late-time spectra share an overall similarity: a smooth continuum plus a Fe bump. However, only the spectrum of 2021csp shows clear line features. If the composition of their ejecta is the same, which is also supported by the similar set of lines in the early spectra, the diversity in the above observable is likely due to a difference in the velocity: higher velocity in SN~2021ckj and slower velocity in SN~2021csp. This can be naturally explained by the above scenario as the ejecta in the polar direction (SN~2021ckj) are faster than in an off-axis direction (SN~2021csp).

Therefore, taking into account the similarities and differences in their photometric and spectroscopic properties, we might be able to understand SNe~2021ckj and 2021csp as follows. They have similar properties of the SN ejecta, CSM and the interaction including an aspherical explosion geometry, but involving different viewing angles. On the other hand, SN~2019hgp has different properties of the ejecta and CSM from those of SNe~2021ckj and 2021csp.


As mentioned in the introduction, there are several proposed scenarios for the progenitors of Type~Icn SNe. The present work provides a new and strong constraint on the nature of the progenitor and explosion: the presence of a high-energy component in the ejecta, which requires the formation of a jet or a collimated outflow. From this point of view, the progenitors of Type~Icn SNe would be more consistent with the following scenarios among those presented in Section 1; a merger of a WR star and a compact object, or a failed/partial explosion of a WR star. In the former case, we expect an enormous diversity in their observational properties originated from the different values of the parameters, such as the different masses of WR stars and companion stars, and different impact parameters. Thus, it might be statistically difficult to have very similar "twins" (SNe~2021ckj and 2021csp) in only three well-observed Type~Icn SNe in the current sample. The latter scenario, which predicts a subrelativistic jet, might explain the observed properties of Type~Icn SNe, even though there are many uncertain processes, e.g., the jet launch mechanism and the fall back processes. For better understanding their progenitors, it is important to study the properties of the ejecta and CSM of Type~Icn SNe with a larger SN sample.

\section{Conclusions} \label{sec:conc}

We have presented photometric and spectroscopic observations of the Type~Icn SN~2021ckj. Its photometric and spectroscopic properties are almost identical to those of SN~2021csp. The photometric evolution is characterised by a high peak brightness ($\sim -20$ mag in the optical bands) and a rapid evolution (rise and above half-maximum times being $\sim 4$ and $\sim 10$ days, respectively, in the \textit{g}/cyan bands). The early spectrum of SN~2021ckj shows narrow emission lines from highly ionized carbon and oxygen lines, while the late-time spectrum is smooth and featureless, except for the Ca~II~triplet line and the iron bump.

The \texttt{TARDIS} modelling of the spectra of SN~2021ckj has clarified that the composition of the SN ejecta is dominated by oxygen and carbon with the iron group elements and the photosperic velocity around the peak is $\sim 10000$ km~s$^{-1}$. The modeling has also implied the need of aspherical SN ejecta.

From the LC modeling, we have found that the ejecta and CSM properties are diverse in Type Icn~SNe 2019hgp, 2021csp and 2021ckj.
The estimated ejecta properties are as follows: SNe~2021ckj and 2021csp must have two components (an aspherical high-energy component and a spherical standard-energy component) with a roughly spherical CSM. On the other hand, SN~2019hgp can be explained by a spherical SN-CSM interaction. In addition, the ejecta of SNe~2021ckj and 2021csp have larger energy per ejecta mass than SN~2019hgp.
These three SNe share a common CSM density distribution, which is similar to those of Type~Ibn SNe. As for the CSM mass, SNe~2021ckj and 2021csp have higher masses than SN~2019hgp, even though both values are within the diversity range of Type~Ibn SNe.

The similarities and differences of the observational properties of SNe~2021csp and 2021ckj can be explained by viewing angle effects of interaction between aspherical ejecta (a collimated high-energy outflow and canonical SN ejecta) and a spherical CSM. We suggest that SN~2021ckj is observed from a direction close to the jet pole, while SN~2021csp from an off-axis direction.

\begin{acknowledgements}
The authors thanks Avishay Gal-Yam and Morgan Fraser for providing the observational data of SNe~2019hgp and 2021csp, and Masaomi Tanaka and Jian Jiang for useful discussions.
This work is based on observations collected at the European Southern Observatory under ESO program IDs 105.20DF (PI: Kuncarayakti) and 1103.D-0328, 106.216C, 108.220C (PI: Inserra; as part of ePESSTO+, the advanced Public ESO Spectroscopic Survey for Transient Objects Survey).
This research made use of \textsc{tardis}, a community-developed software package for spectral synthesis in supernovae \citep{Kerzendorf2014, Kerzendorf2022}. The development of \textsc{tardis} received support from GitHub, the Google Summer of Code initiative, and from ESA's Summer of Code in Space program. \textsc{tardis} is a fiscally sponsored project of NumFOCUS. \textsc{tardis} makes extensive use of Astropy and Pyne.
This research has made use of the NASA/IPAC Infrared Science Archive, which is funded by the National Aeronautics and Space Administration and operated by the California Institute of Technology.

T.N. and H.K. are funded by the Academy of Finland projects 324504 and 328898. T.N. acknowledges the financial support by the mobility program of the Finnish Center for Astronomy with ESO (FINCA).
K.M. acknowledges support from the Japan Society for the Promotion of Science (JSPS) KAKENHI grant JP18H05223, JP20H00174, and JP20H04737. The work is partly supported by the JSPS Open Partnership Bilateral Joint Research Projects between Japan and Finland (K.M. and H.K.; JPJSBP120229923). 
A.P., L.T. are supported by the PRIN-INAF 2022 project "Shedding light on the nature of gap transients: from the observations to the models".
S.M. acknowledges support from the Academy of Finland project 350458.
K.U. acknowledges financial support from Grant-in-Aid for the Japan Society for the Promotion of Science (JSPS) Fellows (22J22705). K.U. also acknowledges financial support from AY2022 DoGS Overseas Travel Support, Kyoto University.
This work was funded by ANID, Millennium Science Initiative, ICN12\_009.
M.G. is supported by the EU Horizon 2020 research and innovation programme under grant agreement No 101004719.
T.E.M.B. acknowledges financial support from the Spanish Ministerio de Ciencia e Innovaci\'on (MCIN), the Agencia Estatal de Investigaci\'on (AEI) 10.13039/501100011033 under the PID2020-115253GA-I00 HOSTFLOWS project, from Centro Superior de Investigaciones Cient\'ificas (CSIC) under the PIE project 20215AT016 and the I-LINK 2021 LINKA20409, and the program Unidad de Excelencia Mar\'ia de Maeztu CEX2020-001058-M.
A.R. acknowledges support from ANID BECAS/DOCTORADO NACIONAL 21202412.
L.G. acknowledges financial support from the Spanish Ministerio de Ciencia e Innovaci\'on (MCIN), the Agencia Estatal de Investigaci\'on (AEI) 10.13039/501100011033, and the European Social Fund (ESF) "Investing in your future" under the 2019 Ram\'on y Cajal program RYC2019-027683-I and the PID2020-115253GA-I00 HOSTFLOWS project, from Centro Superior de Investigaciones Cient\'ificas (CSIC) under the PIE project 20215AT016, and the program Unidad de Excelencia Mar\'ia de Maeztu CEX2020-001058-M.
\end{acknowledgements}

\section*{Data availability}

All the spectroscopic data presented in this paper will be made available on WISeREP. The photometric data are presented in Table~\ref{tab:photo2}.


\bibliographystyle{aa} 
\bibliography{aa.bib}




\end{document}